\documentclass[a4paper]{jpconf}
\usepackage[dvips]{graphicx,color}
\usepackage{epsfig}
\usepackage{amsmath}
\usepackage{amssymb}
\renewcommand{\lsim}{\raise0.3ex\hbox{$\;<$\kern-0.75em\raise-1.1ex\hbox{$\sim\;$}}}
\newcommand{\gsim}{\raise0.3ex\hbox{$\;>$\kern-0.75em\raise-1.1ex\hbox{$\sim\;$}}}

\def\slep{\tilde{l}}

\begin{document}

%%%%% declarations for front matter%%%%%%%%%%%%%%%%%%%%%%%%%%%%%%%%%%%
\title{Long-Lived Slepton in the Coannihilation Region and Measurement of Lepton Flavour Violation at LHC}

\author{S Kaneko$^{1,~2}$, J Sato$^3$, T Shimomura$^4$, O Vives$^4$, M Yamanaka$^3$}

\address{$^1$Instituto F\'isica Corpuscular - C.S.I.C/Universitat de Val\`encia Campus de Paterna, 
        Apt 22085, E46071, Val\`encia, Spain}
\address{$^2$ CFTP, Departamento de F\'isica Instituto Superior T\'ecnico, Avenida Rovisco Pais,1
1049-001 Lisboa, Portugal }
%\ead{satoru@cftp.ist.utl.pt}

\address{$^3$ Department of Physics, Saitama University, Shimo-Okubo, Sakura-ku, Saitama, 338-8570, Japan}

%\ead{joe@phy.saitama-u.ac.jp}
%\ead{masa@krishna.th.phy.saitama-u.ac.jp}

\address{$^4$ Departament de F\'isica Te\`orica and IFIC, Universitat de Val\`encia - CSIC, 
        E46100, Burjassot, Val\`encia, Spain}

%\ead{takashi.shimomura@uv.es}
%\ead{oscar.vives@uv.es}

%%%%% typeset front matter (including abstract) %%%%%%%%%%%%%%%%%%%%%%
\begin{abstract}
When the mass difference between the lightest slepton and the lightest neutralino  
is smaller than the tau mass, the lifetime of the lightest slepton in the constrained 
Minimal Supersymmetric Standard Model (MSSM) increases in many
orders of magnitude with respect to typical lifetimes of other supersymmetric 
particles. In a general MSSM, the lifetime of the lightest slepton is inversely 
proportional to the square of the intergenerational mixing in the slepton mass matrices.
Such a long-lived slepton would produce a distinctive signature at LHC and a 
measurement of its lifetime would be relatively simple. Therefore, the long-lived 
slepton scenario offers an excellent opportunity to study lepton flavour
violation at ATLAS and CMS detectors in the LHC and an improvement of the
leptonic mass insertion bounds by more than five orders of magnitude would be
possible. 
\end{abstract}

%\maketitle

\section{Introduction}
%\subsection{dark matter and LSP}
Cosmological observations have confirmed the existence of the non-baryonic
dark matter. The observed dark matter relic abundance ($\Omega_{DM}h^2 \simeq 0.11$) 
suggests the existence of a stable 
and weakly interacting particle with a mass $100 - 1000$ GeV.
In supersymmetric (SUSY) models with conserved R-parity, the Lightest Supersymmetric 
Particle (LSP), usually the lightest neutralino, is stable and is a perfect 
candidate for the dark matter. In fact, although the dark matter abundance at an
arbitrary point of the SUSY parameter space is usually too large, the observed 
abundance can be reached in regions where the LSP abundance is effectively lowered by annihilation.
One of the mechanisms to lower the dark matter abundance is the so-called
coannihilation  \cite{Griest:1990kh}. In the coannihilation region, the 
Next to Lightest Supersymmetric Particle (NLSP) has a mass nearly degenerate
to the LSP. Then the NLSP and the LSP decouple from thermal bath almost simultaneously
and the LSP annihilates efficiently through collisions with the NLSP. 
The degeneracy required for the coannihilation to occur is generally 
$\delta m/m_{\text{LSP}} <$ a few \%, where 
$\delta m = m_{\text{NLSP}}-m_{\text{LSP}}$. 
%Numerically, this mass difference at a
%fixed $\tan \beta$ value ranges from $\delta m \simeq 10$ GeV for low $M_{1/2}$ to
%practically zero at large $M_{1/2}$, where $M_{1/2}$ is the universal gaugino mass.  

In addition, it was pointed out in \cite{Jittoh:2005pq,Jittoh:2007fr, Jittoh:2008eq} 
that if staus are the NLSP and its mass difference with the LSP is less than
the tau mass,  they destroy $^7$Li/$^7$Be nucleus through the internal conversion during Big-Bang 
Neucleosynthesis (BBN). 
With this small mass difference, staus become long-lived and survive until BBN starts.
Then, they form bound states with nucleus. 
The stau-nucleus bound states decay immediately by virtual exchange of the 
hadronic current.  
In this way, the relic abundance of the light elements is 
lowered effectively and the discrepancy between 
the observed value of $^7$Li/$^7$Be abundance \cite{Ryan:2000zz,Cyburt:2003fe} 
and the predicted value from 
the standard  BBN \cite{Coc:2003ce} with WMAP data 
\cite{Jarosik:2006ib,Dunkley:2008ie} can be solved. 
Thus, a scenario with stau NLSP and neutralino LSP with $\delta m \leq m_\tau$ 
could explain the relic abundance of the light elements as well as the 
abundance of the dark matter. 
%At this point we could ask whether this small
%mass difference  is really possible in the parameter space of the Minimal
%Supersymmetric Standard Model (MSSM) or even the Constrained MSSM (CMSSM) 
%consistently with dark matter constraints. 
%As we will show below, $\delta m \leq m_\tau$ and even smaller values 
%are possible in the CMSSM. 
%However, in this paper we will be more general and ask ``only''  $\delta m \leq m_\tau$. 
%As shown in \cite{Jittoh:2005pq}, if the mass difference, $\delta m$, between neutralino 
%and stau is less than the tau mass, the two-body decay is forbidden and the
%lifetime of stau increases by more than 10 orders of magnitude.

Strictly speaking, the above discussion is true only in the framework of 
the MSSM without intergenerational mixing in the slepton sector.
%as could be,
%for instance, the CMSSM defined at the GUT scale without right-handed
%neutrinos, or neutrino Yukawa couplings. In this case, the NLSP is only a
%combination of left and right-handed staus. 
However, in a general MSSM, we naturally expect some degree of intergenerational mixing
in the sfermion sector. For example, even if we start from a completely universal MSSM at
the GUT scale in the presence of neutrino Yukawa couplings, a small mixing
between different sleptons is generated by renormalization group equations. 
Moreover, there is no fundamental reason to restrict the flavour
structures to the Yukawa couplings and keep the soft mass matrices
universal. In fact, we would expect the same mechanism responsible for the
origin of flavour to generate some flavour mixing in the sfermion sector. 
%As an example, we have flavour symmetries \cite{Ross:2000fn,Raidal:2008jk} where we expect similar
%flavour structures in the Yukawas and the scalar mass matrices. 
In general, we expect that the lightest slepton, $\slep_1$, is not a pure stau, but there is
some mixing with smuon and selectron. In the presence of this small
intergenerational mixing, even with   $\delta m \leq m_\tau$,   
other two-body decay channels like $\slep_1 \rightarrow \tilde{\chi}^0_1 + e(\mu)$, 
are still open \cite{Jittoh:2005pq}. 
In this case, given that the flavour conserving two-body decay 
channel is closed, the slepton lifetime will have a very good sensitivity to 
small Lepton Flavour Violation (LFV) parameters. The lifetime will be inversely proportional 
to the sfermion mixing until two-body decay width becomes comparable to three or 
four-body decay widths. Thus, it is worthwhile to study the lifetime of the 
sleptons in small $\delta m $ case to obtain information on LFV parameters.
Several papers have studied the possibility of measuring LFV at
colliders
\cite{Agashe:1999bm,Deppisch:2003wt,Deppisch:2004pc,Bartl:2005yy,Ibarra:2006sz,HohenwarterSodek:2007az,Hirsch:2008dy,Hirsch:2008gh}.
However, in these papers the LFV decays are always
subdominant with small branching ratios and is never possible to reach 
the level of sensitivity we reach in our scenario. 
Only Ref.~\cite{Ibarra:2006sz}
considers LFV in  $e^+ e^-$ colliders through the slepton production in a 
long-lived stau scenario in a gauge-mediated model and as in the other works
they are not sensitive either to the presence of small lepton flavour 
violation that we consider in this paper. 

Long-lived charged-particles are very interesting since they provide a clear
experimental signature at the LHC \cite{Hamaguchi:2006vu,tarem:2008,tarem:2008ph}. 
In Ref.~\cite{Ishiwata:2008tp}, it is concluded
that even if the lifetime of the decaying particle is much longer than the
size of the detector, some decays always take place inside the detector and 
it is possible to measure lifetimes as long as $10^{-5} - 10^{-3}$ seconds
in a particular gauge mediation scenario. 

%\subsection{What we do}
%In this paper, we study LFV in the coannihilation region of the MSSM with 
%neutralino LSP and slepton NLSP with a mass difference $\delta m \leq m_\tau$. 
%The paper is organized as follows. In Sec. 2, we study the case of the MSSM
%without flavour mixing in the slepton sector. Here, we show that a sizable
%part of the coannihilation region at large $M_{1/2}$ has  
%$\delta m \leq m_\tau$ region 
%satisfying all the experimental constraints and we calculate the lifetime of the NLSP.  
%Then, in Sec. 3, we introduce a source of lepton flavour violation in the
%slepton mass matrices and we recalculate the lifetimes in terms of the LFV
%parameters. In Sec. 4, we discuss the expected phenomenology of this scenario 
%at the LHC and how LHC data could be used to measure small LFV parameters.
%Finally in section 5, we present our summary and discussion.                         

\section{Long-lived Stau in MSSM} \label{long-lived-mssm}
We start the analysis looking for allowed regions in the Constrained MSSM (CMSSM) parameter space where the difference, $\delta m$, 
between the lightest neutralino mass, $m_{\tilde{\chi}^0_1}$, and the lighter stau mass, $m_{\tilde{\tau}_1}$, is smaller 
than the tau mass, $m_\tau$. In this section, we consider the case of a CMSSM defined at the GUT scale without 
neutrino Yukawa couplings. Then, $\tilde{\tau}_1$ is equal to a combination of only the right and left-handed stau. 
The CMSSM is parametrized by $4$ parameters (the universal gaugino masses, $M_{1/2}$, 
the universal scalar masses, $m_0$, the universal trilinear couplings, $A_0$, at the GUT scale and the ratio of vacuum 
expectation values of two Higgses, $\tan\beta$) and one sign (of $\mu$).
In our numerical analysis, we use SPheno \cite{Porod:2003um} and micrOMEGAs \cite{Belanger:2008sj} to obtain mass spectrum at electroweak
scale and relic dark matter abundance.

First, we list the experimental constraints used in our analysis. The main constraint, apart from 
direct limits on SUSY masses, comes from the relic density of the dark matter abundance reported by WMAP collaboration.
Although the latest data is obtained in \cite{Dunkley:2008ie}, we use more conservative range in this paper, 
$0.08 < \Omega_{\mathrm{DM}}h^2 < 0.14$, 
%\begin{align}
%0.08 < \Omega_{\mathrm{DM}}h^2 < 0.14, \label{conserv_range}
%\end{align}
but the results would be basically the same as the one with a narrower range.

A second constraint comes from the anomalous magnetic moment of the muon,
$a_\mu=(g-2)_\mu /2$, that has been measured precisely in the experiments 
at BNL \cite{Bennett:2006fi} 
The difference between the experiment and the standard model prediction is given \cite{Davier:2007ua} as 
$\Delta a_\mu=a_\mu^{(\mathrm{exp})} - a_\mu^{(SM)}=(27.5 \pm 8.4) \times 10^{-10}$, 
%\begin{align}
% \Delta a_\mu=a_\mu^{(\mathrm{exp})} - a_\mu^{(SM)}=(27.5 \pm 8.4) \times 10^{-10}
%\end{align}
which corresponds to a difference of $3.3$ standard deviations. These results seem to require a positive and sizable SUSY contributions 
to $a_\mu$, although it is still not conclusive. In the
following, in all our plots we will show the regions favored by measurement of $a_\mu$ 
at a given confidence level, but we will not use it as a constraint to exclude
the different points. The bound on Br$(b \rightarrow s \gamma)$ is also taken into account,
but as we will see, in the region of small $\delta m$ and correct dark matter abundance, $M_{1/2} \ge 700$ GeV.
Therefore the numerical prediction we obtained for  Br$(b \rightarrow s \gamma)$ is always very close to the SM predictions.

\begin{figure}[h]
\begin{center}
\begin{tabular}{ccc}
\hspace{-10mm}\includegraphics[height=57mm]{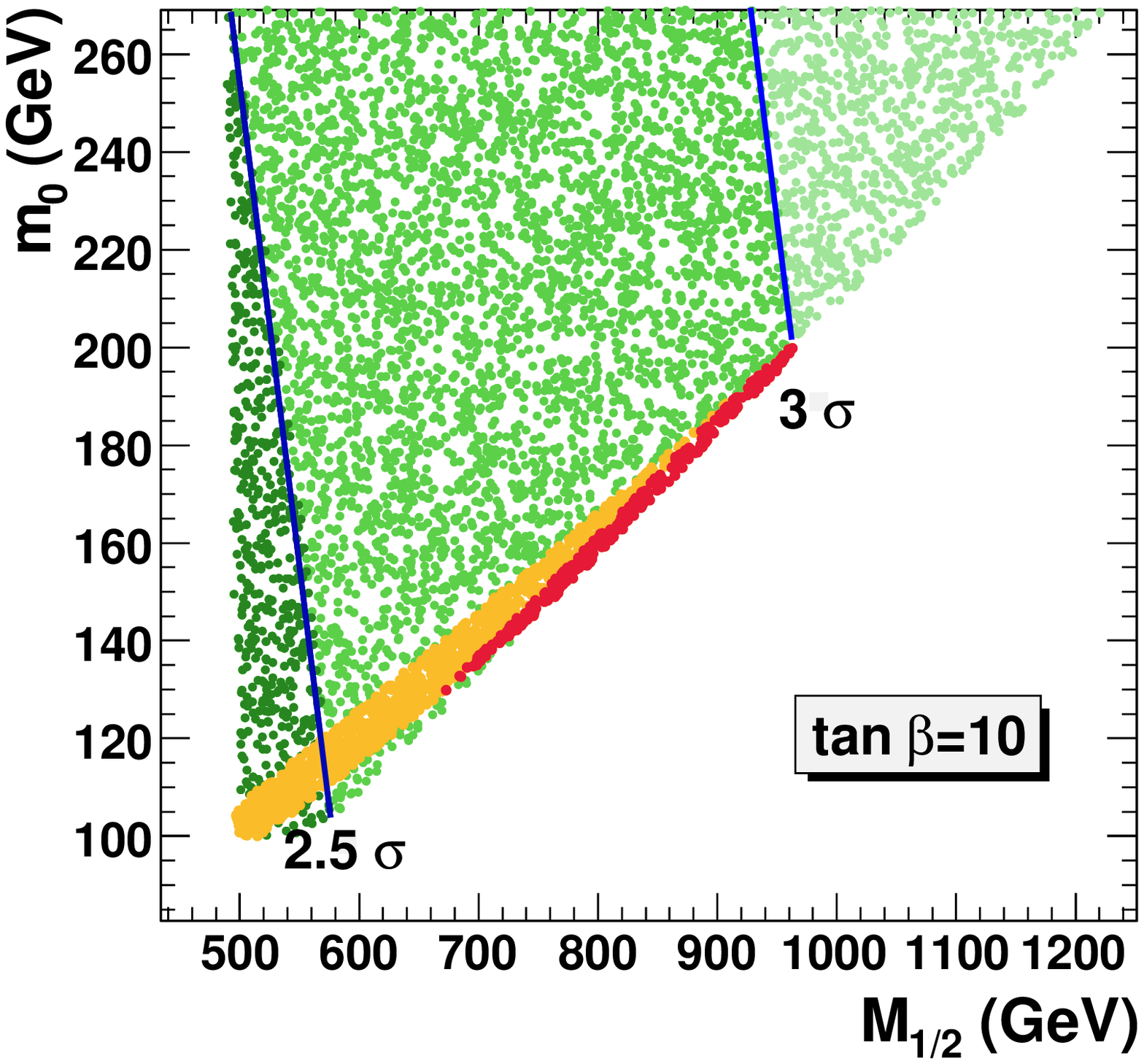} & 
\hspace{-10mm}\includegraphics[height=57mm]{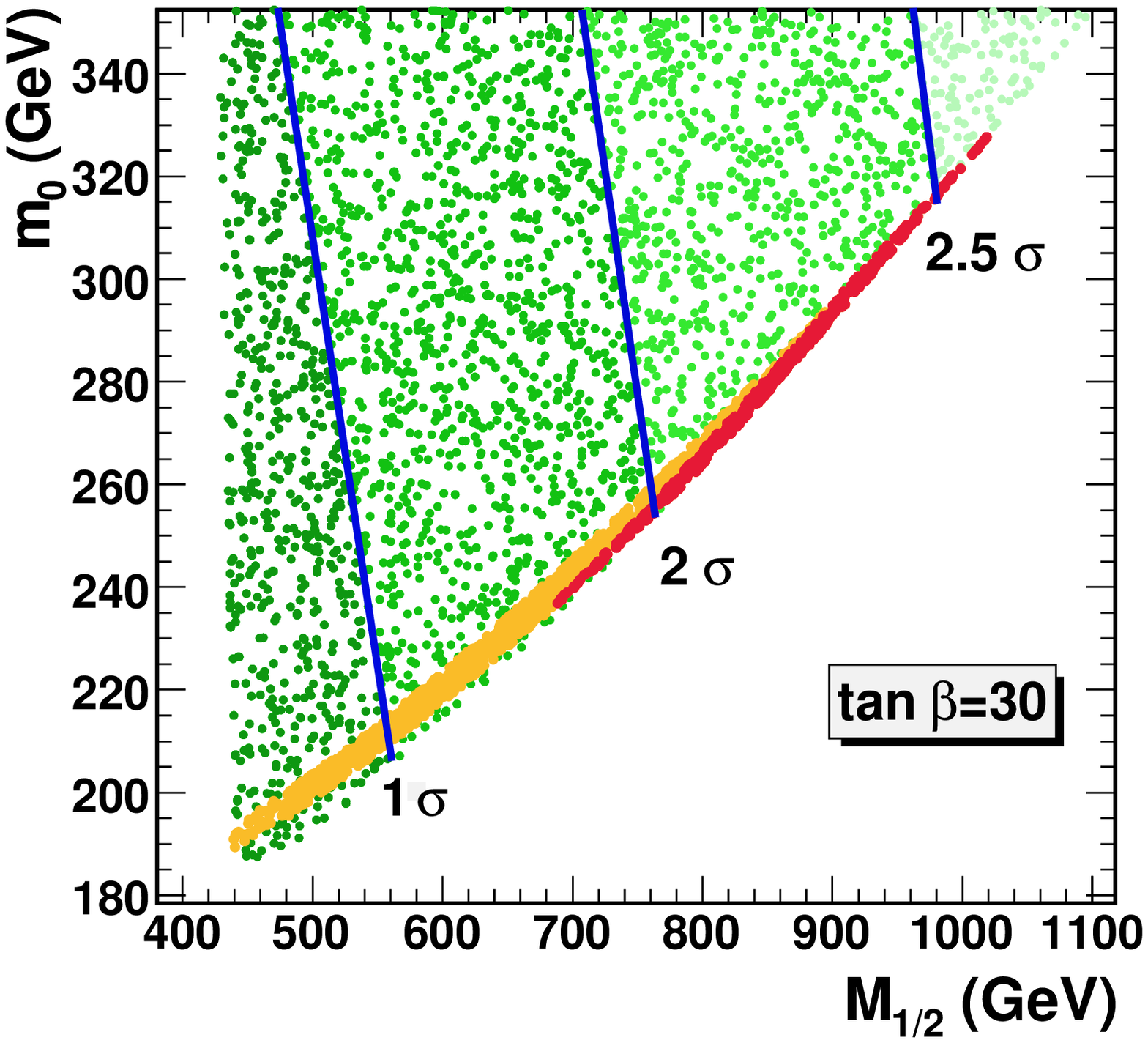} & 
\hspace{-10mm}\includegraphics[height=57mm]{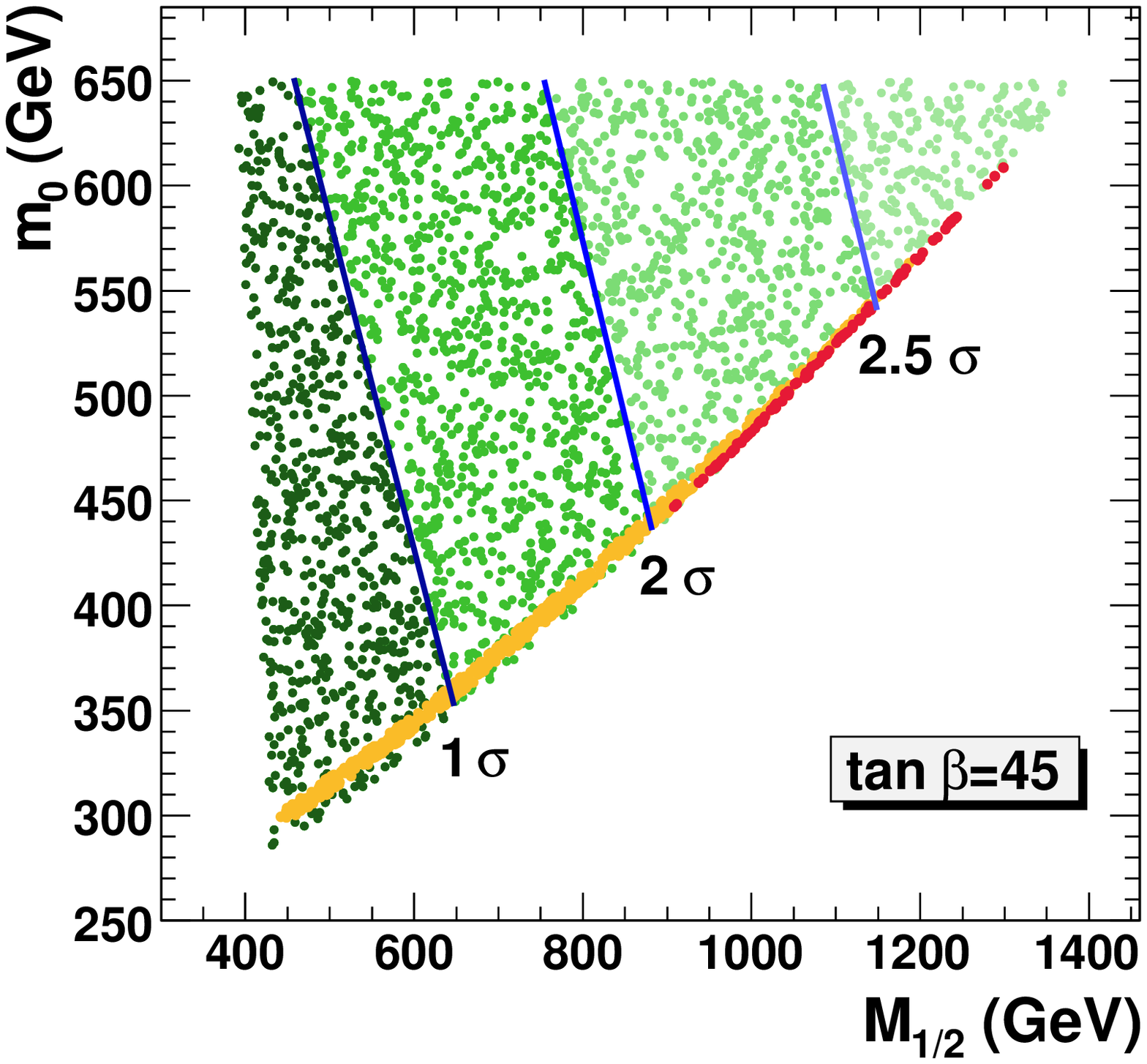}
%\multicolumn{2}{c}{\includegraphics[height=70mm]{figures/tgb45.eps}}
\end{tabular}
\end{center}
%\vspace{-1cm} 
\caption{The allowed parameter regions in $M_{1/2} - m_0$ plane fixing $A_0=600$ GeV. $\tan\beta$ is varied $10$, $30$, $45$ shown 
in each figure. The red (dark) narrow band is consistent region with dark matter abundance and $\delta m < m_\tau$ and 
the yellow (light) narrow band is that with $\delta m > m_\tau$. The green regions are inconsistent with the dark matter abundance, and 
the white area is stau LSP region and excluded. The favored regions of the muon anomalous magnetic moment 
at $1\sigma$, $2\sigma$, $2.5\sigma$ and $3\sigma$ confidence level are indicated by solid lines.}
\label{mh_m0}
\end{figure}
In Fig.~\ref{mh_m0}, we show the allowed parameter regions in the $M_{1/2}$ -- 
$m_0$ plane for $\tan\beta=10,~30,~45$ and $A_0=600$ GeV. The sign of $\mu$ is 
taken to be positive in all figures as required by the $b\to s \gamma$ and
$a_\mu$ constraints. In these figures,  the solid lines indicate the 
different confidence level  regions for the $a_\mu$
constraint. The yellow (light) narrow band is the region with correct dark
matter  abundance corresponding to the coannihilation region and with 
$\delta m > m_\tau$. The red (dark) band in the coannihilation region represents 
the points with correct dark matter abundance and $\delta m < m_\tau$. 
The green regions are not consistent with the dark matter abundance while the 
white area is the stau LSP region and therefore is excluded.  
Note that in models where the observed dark matter abundance is accounted for other dark matter components besides 
the neutralino, the small $\delta m$ region could be extended to 
the green regions below the yellow band.  
Notice that choosing a positive and large $ A_0~(\lsim 3 m_0)$, the renormalization group evolved
value $A (M_W)$ is reduced with respect to the cases of zero or negative
$A_0$. Therefore the left-right entry in the stau mass matrix, $A - \mu \tan \beta$, 
is also reduced. 
Then the region of neutralino LSP, for a fixed value of the neutralino mass, 
$m_{\tilde{\chi}^0_1} ~(\simeq 0.4 M_{1/2})$, corresponds to smaller values of $m_0$ 
for positive $A_0$ than for zero or negative $A_0$.
Therefore this implies slightly lighter spectrum and 
larger contributions to $a_\mu$. 
%Anyway, the main features of the
%plots are maintained for different values of $A_0$ with only this slightly
%increase in the masses. 
From these plots we can see that the interesting region 
of correct value of dark matter relic abundance and $\delta m < m_\tau$ corresponds always to
relatively large values of $M_{1/2}$. 
This is due to the fact that the 
coannihilation cross section decreases at larger SUSY masses \cite{Ellis:1999mm} 
and for sufficiently large $M_{1/2}$ the cross section is too small to produce a correct dark matter 
abundance, even for $m_{\tilde{\tau}_1} \simeq m_{{\chi}^0_1}$. 

Comparing the different $\tan \beta$ values, we can see that
at low $\tan \beta$ the long-lived stau region corresponds to smaller values
of $m_0\sim (120,200)$ GeV, although it can only reach the $a_\mu$ favored
region at 3 $\sigma$. For $\tan\beta =30$, the long-lived stau region
corresponds to similar values of $M_{1/2}$, but the required values of $m_0$
are nearly a factor two larger. However, in this case the larger value of
$\tan \beta$ allows this region to reach the $a_\mu$ favored region
at the $2\sigma$ level. At $\tan\beta =45$ the long-lived stau region
corresponds to larger values of both $M_{1/2}$ and $m_0$ and it can only reach
the $a_\mu$ favored region at 2.5 $\sigma$. This behavior of the allowed 
regions can be understood as follows.
Since the SUSY contribution to $a_\mu$ is proportional to $\tan\beta/m^2_{\tilde{\chi}^0_1}$, it is 
small when $\tan\beta=10$. But it is also suppressed at $\tan\beta > 30$ because the mass of the neutralino 
becomes heavier for large $\tan\beta$. With fixed $A_0$, the neutralino mass is determined mainly by $M_{1/2}$ and 
the increase of the neutralino mass according to the increase of $\tan\beta$ is seen in Fig.~\ref{mh_m0}.

The decay rates of the NLSP stau into two, three and four bodies are approximately given in \cite{Kaneko:2008re}.
%\begin{align}
% \Gamma_{\mathrm{2-body}} &=\frac{g_2^2 \tan^2\theta_W}{2\pi m_{\tilde{\tau}_1}} \delta m \sqrt{(\delta m)^2 - m_\tau^2}, \label{lfc_2_body}\\
% \Gamma_{\mathrm{3-body}} &=\frac{g_2^2 G_F^2 f_\pi^2 \cos^2\theta_c \tan^2\theta_W }{30 (2\pi)^3 m_{\tilde{\tau}_1} m_\tau^2}~
%  \delta m \big( (\delta m)^2 - m_\pi^2 \big)^{5/2}, \\
% \Gamma_{\mathrm{4-body}} &=\frac{2}{3}\frac{ g_2^2 G_F^2 \tan^2\theta_W}{5^3 (2\pi)^5 m_{\tilde{\tau}_1} m^2_\tau} 
%  \delta m \big( (\delta m)^2 - m_l^2 \big)^{5/2} \big( 2(\delta m)^2 - 23 m_l^2 \big),
%\end{align}
%where the notations and the relevant Lagrangian, the exact decay rates are given in \cite{Kaneko:2008re}. 
Notice that when $\delta m \gsim m_\tau$ the two-body decay is open and the
lifetime of the stau, $\tau_{\tilde{\tau}_1}$, is $ \lsim 10^{-22}$ sec. However, for $\delta m
\lsim m_\tau$ this decay is closed and the three or four-body decays are
suppressed at least by an additional 
$(\delta m)^4 G_F^2 (f_\pi/m_\tau)^2 1/(30 (2\pi)^2) \simeq 10^{-13}$ with $\delta m \sim 2$ GeV. Therefore
the stau becomes long-lived and the phenomenology of the MSSM changes
dramatically.
%\footnote{An interesting question in the long-lived stau region is the possibility to 
%detect a long-lived stau at neutrino telescopes through inelastic
%scattering of high-energy neutrinos. However the expected number of events in
%our scenario is always small and the neutrino telescope will not put 
%a strong constraint \cite{Canadas:2008}.}

\section{LFV and Long-lived Slepton}
In the previous section, we have seen that $\delta m \leq m_\tau$ is indeed
possible in a CMSSM without LFVs. As was shown in \cite{Jittoh:2005pq}, the lifetime of the NLSP stau increases by many orders of magnitude. 
However, in realistic models, we expect a certain degree of intergenerational mixing to be present in the
slepton sector. Once a new source of LFVs is introduced, the NLSP two-body 
decay channels into electron and/or muon is open again. In this case, the 
lifetime is inversely proportional to the square of the mixing of selectron and smuon with stau and  
therefore the measurement of the lifetime shows a strong sensitivity to LFV parameters. 
In this section, we show the dependence of the lifetime of the lightest
slepton on the right-handed and the left-handed slepton mixings. All the numerical results 
of lifetimes and figures presented below are calculated using the exact formula given in \cite{Kaneko:2008re}.

To understand the dependence of lifetimes on LFV parameters, 
it is convenient to introduce the so-called Mass Insertions (MI), $(\delta_{RR/LL}^e)_{\alpha \beta}$, 
defined in the same way as \cite{Kaneko:2008re}.
%\begin{align}
% (\delta^e_{RR/LL})_{\alpha \beta} = \frac{\Delta {M^{e~2}_{RR/LL}}_{\alpha \beta}}{M^e_{R/L \alpha} M^e_{R/L \beta}}, \label{MI_param}
%\end{align}
%where $\alpha,~\beta$ denote the lepton flavours. 
%$M^e_{R/L \alpha}$ and $M^e_{R/L \beta}$ are diagonal elements of the slepton mass matrix, and 
%$\Delta {M^{e}_{RR/LL \alpha \beta}}$ are off-diagonal elements we introduced. 
In terms of these mass insertions, the two-body decay rate is approximately 
given by 
\begin{align}
 \Gamma_{\mathrm{2-body}} &= \frac{g_2^2}{2\pi m_{\tilde{\tau}_1}}(\delta m)^2 (|g_{1\alpha 1}^L|^2+|g_{1\alpha 1}^R|^2), \label{lfv_2_body}
\end{align}
where $\alpha=e,~\mu$. 
$g^{L,R}_{1\alpha 1}$ can be approximated in the mass insertion as shown.
%\begin{figure}[t]
%\begin{center}
%\begin{tabular}{ccc}
%\includegraphics[height=42mm]{figures/MI_right.eps} & ~~~ &
%\includegraphics[height=42mm]{figures/MI_left.eps}
%\end{tabular}
%\caption{MI Feynman diagrams of two-body slepton decay: The diagram on the left side 
%depicts two-body decay in the presence of $\delta^e_{RR}$ and the diagram on the right side 
%in the presence of $\delta^e_{LL}$. The circle-crosses represent mass insertions for flavor and left-right chirality 
%changes. Propagators for intermediate states are shown below corresponding scalar lines.}
%\label{MI_diagrams}
%\end{center}
%\end{figure} 
In the case of right-handed slepton mixing, we have, 
\begin{align}
 g^L_{1\alpha 1}\simeq 0,\quad 
 g^R_{1\alpha 1}\simeq \tan\theta_W 
  \frac{M^e_{R\tau} M^e_{R\alpha}}{M^{e~2}_{R\tau}-M^{e~2}_{R\alpha} }(\delta^e_{RR})_{\alpha \tau}, \label{eR_g}
\end{align}
while in the left-handed slepton mixing case, these couplings are given
\begin{align}
 g^L_{1\alpha 1}\simeq \frac{1}{2}\tan\theta_W 
\frac{m_\tau( A_0-\mu \tan\beta )}{ M^{e~2}_{R \tau} - {M^{e~2}_{L\tau}} } 
\frac{M^e_{L \alpha} M^e_{R\tau}}{M^{e~2}_{R\tau} - M^{e~2}_{L\alpha}}(\delta^e_{LL})_{\alpha \tau}, 
\quad g^R_{1\alpha 1}\simeq 0. \label{eL_g}
\end{align}

\begin{figure}[t]
\begin{center}
\begin{tabular}{cc}
\includegraphics[height=55mm]{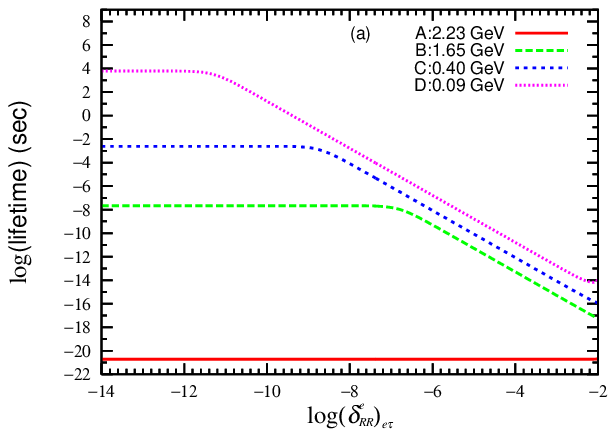} &
\includegraphics[height=55mm]{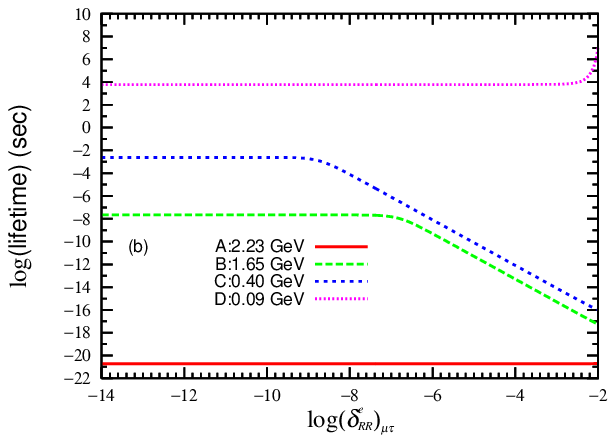}
\end{tabular}
\caption{The lifetime of the lightest slepton as a function of $\delta^e_{RR}$. The left panel,(a), is the lifetime of the lightest slepton with 
the right-handed selectron and stau mixing, 
and the right panel, (b), is the one with the right-handed smuon and stau mixing. In both panel, the solid (red),
dashed (green), dotted (blue), dotted-dashed (pink) line correspond to the point $A,~B,~C,~D$ of Table. 1, respectively. 
$m_0=260$ GeV, $A_0=600$ GeV and $M_{1/2}$ is varied.}
\label{lifetime_right}
\end{center}
\end{figure}

To analyze the effects of the presence of a non-vanishing leptonic mass
insertion on the NLSP lifetime, we choose four points with different mass
differences as shown in Table~\ref{param_tab}.
\begin{table}[t]
\caption{Table of the mass difference and the lightest slepton, neutralino masses. 
$m_0$, $A_0$ and $\tan\beta$ are fixed to $260$ GeV, $600$ GeV and $30$, respectively.
The values of neutralino abundance and $a_\mu$ are shown for the reference.}
\begin{center} 
\begin{tabular}{cccccc}\br
 ~~No.~~ & \hspace{2mm} $\delta m$ (GeV) \hspace{2mm} & \hspace{2mm} $m_{\tilde{\chi}_1^0}$ (GeV) \hspace{2mm} & \hspace{2mm} $m_{\tilde{l}_1}$
 (GeV) \hspace{2mm} & \hspace{2mm} $\Omega_{\tilde{\chi}^0_1} h^2$ \hspace{2mm} & \hspace{2mm} $a_\mu$ ($\times 10^{-10}$) \hspace{2mm} \\ \mr
  A  & 2.227  & 323.1549 & 325.3817 & 0.110 & 10.32 \\
  B  & 1.650  & 325.5601 & 326.2147 & 0.102 & 10.25 \\
  C  & 0.407  & 327.6294 & 328.0365 & 0.085 & 10.09 \\
  D  & 0.092  & 328.4060 & 328.4981 & 0.081 & 10.06 \\ \br
\end{tabular}
\end{center}
\label{param_tab}
\end{table}
In Fig.~\ref{lifetime_right} (a),  we show the lifetime of the lightest slepton,
$\tau_{\tilde{l}_1}$, as a function of $(\delta_{RR}^e)_{e\tau}$, encoding the right-handed
selectron-stau mixing, that we vary from $10^{-10}$ to $10^{-2}$. 
We can see that the lifetime for $\delta m > m_\tau$ (case A in Table. \ref{param_tab}) does not change,
because the decay of slepton to tau and neutralino is always the 
dominant decay mode and the lifetime is insensitive to $\delta^e_{RR}~(\leq
10^{-2})$. On the other hand, for $\delta m < m_\tau$, the lifetime grows more than $13$
orders of magnitude in the limit $(\delta^e_{RR})_{e\tau} \rightarrow 0$, 
where the three- or four-body decay processes are dominant. Then,  the two-body decay into 
$\tau$ and $\tilde{\chi}^0_1$ is forbidden but those into $e$ (or $\mu$ for 
$(\delta_{RR}^e)_{\mu\tau}\neq 0$) and
$\tilde{\chi}^0_1$ are allowed through LFV couplings. The lifetime
decreases proportionally to $|(\delta^e_{RR})_{e\tau}|^{-2}$ when the two-body 
decay dominates total decay width. For values of $\delta^e >10^{-2}$ the mass of the lightest slepton would be substantially changed by 
this large off-diagonal entry and this would reduce the mass difference
changing the simple $|(\delta^e_{RR})_{e\tau}|^{-2}$ proportionality. This can be 
seen as an increase of the lifetime in the case D at $(\delta^e_{RR})_{\mu\tau} \simeq 10^{-2}$ in Fig.~\ref{lifetime_right} (b).
Although $\delta^e_{RR} > 10^{-2}$ is still allowed by experiments,
% as shown in Table \ref{Table1},
we concentrate our discussion on $\delta^e_{RR} \leq 10^{-2}$ to show the 
sensitivity of this process to small $\delta^e_{RR}$s.
In fact, from these figures we can see that the lifetime indeed has very good 
sensitivity to small $\delta^e$s in this scenario. 
Comparing the bounds on the mass insertions given in \cite{Kaneko:2008re} with the figures we can see that the 
sensitivity obtained in the measurement of the NLSP lifetime can not 
be reached by indirect experiments as $\tau \rightarrow \mu~\gamma$ and $\tau \rightarrow e~\gamma$, etc. 
Plot (b) in Fig.~\ref{lifetime_right} corresponds to the 
right-handed smuon-stau mixing, $(\delta_{RR}^e)_{\mu\tau}$. 
The lifetime is constant for $\delta m = 0.09$ GeV in small
$(\delta^e_{RR})_{\mu\tau}$ as the two-body decay in muon is still forbidden.
For larger $\delta m$, we observe again the same dependence on
$(\delta_{RR}^e)_{\mu\tau}$. 

\begin{figure}[t]
\begin{center}
\begin{tabular}{cc}
\includegraphics[height=55mm]{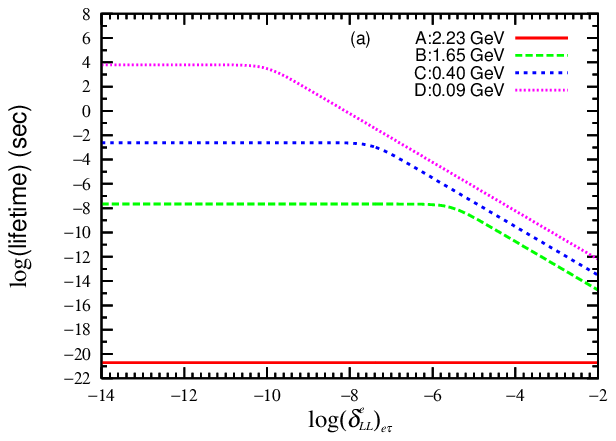} & 
\includegraphics[height=55mm]{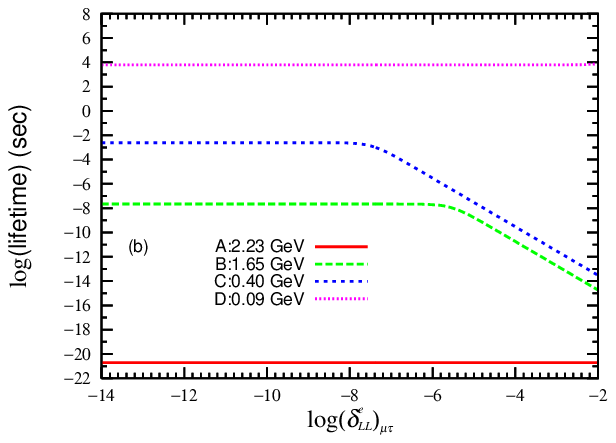}
\end{tabular}
\caption{The lifetime of the lightest slepton as a function of $\delta^e_{LL}$. The lines and the parameters are the same as 
Fig.\ref{lifetime_right}.}
\label{lifetime_left}
\end{center}
\end{figure}

%\begin{table}[t]
%\begin{center}
% \begin{tabular}{|c||c|c|c||c|c|c||c|c|c|}
% \hline
%
%&\multicolumn{3}{c||}{$\tan\beta=10$} &
%\multicolumn{3}{c||}{$\tan\beta=30$ } &
%\multicolumn{3}{c|}{$\tan\beta=45$}\\
%\hline
%& $\delta^e_{e\mu}$ & $\delta^e_{e\tau}$ & $\delta^e_{\mu\tau}$& $\delta^e_{e\mu}$ &
%$\delta^e_{e\tau}$ & $\delta^e_{\mu\tau}$& $\delta^e_{e\mu}$ & $\delta^e_{e\tau}$ &
%$\delta^e_{\mu\tau}$ \\
%\hline
%$LL$ & 0.0014 & 0.33 & 0.21 & 0.00047 & 0.10 & 0.068 & 0.00030 & 0.067
%& 0.043\\
%$RR$ & 0.0060 & 1.7 & 1.1 & 0.0019 & 0.44 & 0.29 & 0.0012 & 0.28 & 0.18\\
%%~$LR/RL$~ & ~$7.8\times10^{-6}$~ & ~0.030~ & ~0.019~ &
%%~$7.7\times10^{-6}$~ & ~0.029~ & ~0.019~ & ~$7.6\times10^{-6}$~ &
%%~0.029~ &
%%  ~0.019~ \\
%\hline
%\end{tabular}
%\caption{\label{Table1} Mass insertion bounds for a typical point in the
%  long-lived stau coannihilation region with $m_0=267$~GeV,
%  $M_{1/2}=750$~GeV, $A_0=600$~GeV and different values of $\tan \beta$
%\cite{Calibbi:2008qt,Masina:2002mv}.
%These bounds are obtained from the experimental limits
%($BR(\mu\to e\gamma)<1.2\times10^{-11},BR(\tau\to
%e\gamma)<1.1\times10^{-7},BR(\tau\to \mu\gamma)<0.45\times10^{-7}$). }
%\end{center}
%\end{table}
Figs.~\ref{lifetime_left} (a) and (b) show the lifetime of the lightest 
slepton as a function of  $(\delta_{LL}^e)_{e\tau}$ and  $(\delta_{LL}^e)_{\mu\tau}$, 
respectively. The different curves correspond to the same mass differences used
in Fig.~\ref{lifetime_right}. The dependence on these mass insertions in these 
figures is completely analogous to the behavior observed in 
Fig.~\ref{lifetime_right}. In fact, both figures would be identical if we
replace $(\delta_{LL}^e)$ by $10 \times (\delta_{RR}^e)$.
This is due to the fact that in this case, the lightest slepton decays into the electron 
or the muon through the left-handed stau component. The mixing of right and left-handed staus,
Eq.~\eqref{eL_g}, is proportional to 
$m_\tau(A_0 - \mu\tan\beta)/( M^{e~2}_{R\tau} - {M^{e~2}_{L\tau}} )$. In the region of
parameter space we are considering  this factor is approximately 
$0.1$. Thus, we can see
that, also in the case of $(\delta_{LL}^e)$, the lifetime is sensitive to the
presence of very small lepton flavour violating couplings and therefore 
to the presence of mass insertions orders of magnitude
smaller than the present bounds.

\section{LHC Phenomenology}
In this section, we discuss the expected phenomenology at LHC experiments, focusing mainly on 
the ATLAS detector \cite{:2008zz,atlashp_url},
of the long-lived slepton scenario. 
The lightest slepton is the NLSP, and therefore a large number of sleptons is expected to be produced via cascade 
decays of heavier SUSY particles. When the long-lived sleptons have lifetimes between $10^{-5}$ and $10^{-11}$ sec., 
we will have a chance to observe decays of the slepton 
inside the detector.
%The ATLAS detector, whose dimensions are $25$ m in hight and $44$ m in length, consists of inner detector, calorimeter 
%and muon detector \cite{:2008zz}. 
%The inner detector surrounds the collision point and is contained within a cylindrical envelop of length 
%$\pm 3.51$ m and of radius $1.15$ m. 
%In the inner detector, the pixel detector, the semiconductor tracker (SCT) and the transition radiation tracker (TRT) 
%are installed to measure momenta and decay vertices of charged-particles. 
%The pixel detector is placed around the collision point and consists of three barrels at average radii of $5$ cm, 
%$9$ cm and $12$ cm, and three disks on each side, between radii of $9$ and $15$ cm. The SCT surrounds the pixel detector and 
%consists of four barrels at radii of $30$, $37$, $44$ and $52$ cm. The TRT is installed on outside of the SCT. It covers radial 
%range from $56$ to $107$ cm. 
%Surrounding the inner detector, hadronic and electromagnetic calorimeters are placed over $4$ m radius and $8.4$ m length. 
%The calorimeters are surrounded by the muon spectrometer which covers the ATLAS detector. 
In the following, we show that, in the 
relevant region of SUSY parameter space,  
we can see decays of sleptons with the lifetime up to $10^{-5}$ seconds within the ATLAS detector. 
Although we discuss only the ATLAS detector, analysis is similar to the CMS detector \cite{:2008zzk}.

First, we estimate the number of lightest sleptons produced at LHC. In our
scenario, the lightest slepton is mainly a right-handed stau and the lightest
neutralino is mainly bino. The spectrum that we consider here 
corresponds approximately  to the post-WMAP SUSY benchmark point $J^\prime$ 
proposed in \cite{Battaglia:2001zp,Battaglia:2003ab} with a slightly
different $\tan \beta$ value ($\tan \beta= 35$ versus $\tan \beta=30$ in our
case).  Then, the right-handed 
staus are mainly produced via decays 
of first two generation left-handed squarks, 
%(number in parenthesis corresponds to branching ratios in each decay \cite{Battaglia:2001zp,Battaglia:2003ab})
%\begin{align}
% \tilde{q}_L &\rightarrow \tilde{\chi}^\pm_1 + q~~(0.63),\\
%  &~~~~~ \tilde{\chi}^\pm_1 \rightarrow \tilde{l}_1 + \nu_\tau~~(0.64),~~\tilde{\nu}_\tau + \tau~~(0.27), \\
%  &~~~~~~~~~~~~~ \tilde{\nu}_\tau \rightarrow \tilde{l}_1 + W~~(0.82), \\
% \tilde{q}_L &\rightarrow \tilde{\chi}^0_2 + q~~(0.36), \\
%  &~~~~~ \tilde{\chi}^0_2 \rightarrow \tilde{l}_1 + \tau~~(0.66),~~\tilde{\nu}_\tau + \nu_\tau~~(0.25),
%\end{align}
%where $q$, $\tilde{\chi}^\pm_1$ and $\tilde{\chi}^0_2$ denote the SM quarks, the lighter 
%charginos and the second lightest neutralinos and $\nu_\tau$, $W$ and $\tilde{\nu}_\tau$ denote 
%tau neutrino, weak boson and tau sneutrino, respectively. 
The 3rd generation squarks have many different decay chains. 
%For example
%\begin{align}
% \tilde{t}_{1,2} &\rightarrow \tilde{\chi}^\pm_1 + b,~~(0.20,~0.25) \\
%  &~~~~~ \tilde{\chi}^\pm_1 \rightarrow \tilde{l}_1 + \nu_\tau, \\
% \tilde{t}_{1,2} &\rightarrow \tilde{\chi}^0_2 + t,~~(0.10,~0.10) \\ 
%  &~~~~~ \tilde{\chi}^0_2 \rightarrow \tilde{l}_1 + \nu_\tau, \\
% \tilde{b}_{1,2} &\rightarrow \tilde{\chi}^\pm_1 + t,~~(0.36,~0.12) \\
%  &~~~~~ \tilde{\chi}^\pm_1 \rightarrow \tilde{l}_1 + \nu_\tau,\\
% \tilde{b}_{1,2} &\rightarrow \tilde{\chi}^0_2+b,~~(0.20,~0.10)\\
%  &~~~~~ \tilde{\chi}^0_2 \rightarrow \tilde{l}_1 + \nu_\tau.
%\end{align}
Total branching ratios for $\tilde{l}_1$ production are $0.86$ in $\tilde{q}_L$ decay, $0.72/0.90$ 
for $\tilde{t}_1/\tilde{t}_2$ and $0.87/0.67$ for $\tilde{b}_1/\tilde{b}_2$ at $\tan\beta=35$
\cite{Battaglia:2001zp,Battaglia:2003ab,diff_tanb}.
On the other hand, the right-handed squarks, couple only to bino and higssino, having small Yukawa couplings, 
decay almost completely to the lightest neutralino and quarks. 
Therefore branching ratios of them into $\tilde{l}_1$ and quarks/leptons are negligible. 
The total cross section of SUSY pair production in our scenario is given as 
$\sigma_{\mathrm{SUSY}}=130~\mathrm{fb}$ in \cite{Skands:2001it}\footnote{Taking into account
that the production cross section is almost identical for different values of
$\tan \beta$, we use the cross section for point $P_2 \simeq J^\prime$ 
in \cite{Skands:2001it}},
%\begin{align}
% \sigma_{\mathrm{SUSY}}=130~\mathrm{fb}.
%\end{align}
When we assume the integrated luminosity, $\mathcal{L_{\mathrm{int}}}=30~\mathrm{fb}^{-1}$, 
%which corresponds to three year of data taking with the luminosity, $\mathcal{L}= 10^{33}$ cm$^{-2}$ s$^{-1}$, 
the total number of SUSY pairs is $3900$.
Then, squarks are pair-produced with equal probability for left and right-handed squarks, thus, 
the number of lightest sleptons produced, $N_{\tilde{l}_1}$, is estimated as $N_{\tilde{l}_1} \simeq 4290$.
%\begin{align}
% N_{\tilde{l}_1} \simeq 4290.
%\end{align}
%Therefore we would have $4290$ long-lived sleptons produced in ATLAS that could decay inside the detector depending 
%on the lifetime. 
%The decay probability is given 
%\begin{align}
% P_{\mathrm{dec}}(l)=1-\exp\left(-\frac{l}{\beta\gamma c \tau_{\tilde{l}_1}}\right), \label{prob_dec}
%\end{align}
%where $\beta=v/c$ and $\gamma=1/\sqrt{1-\beta^2}$ with $c$ the speed of the light, $l$ is the distance between production and decay point. 
To obtain the expected number of decays, we need to know the distribution of $\beta\gamma$ where $\beta=v/c$ and $\gamma=1/\sqrt{1-\beta^2}$ 
with $c$ the speed of the light. This would require a full analysis with 
Monte-Carlo simulation and it is beyond the scope of this paper. For the spectrum we consider, the typical momentum of the 
lightest slepton is expected to be between $500$ and $900$ GeV. $\beta\gamma$ corresponding to this range of momenta is 
$ 1.53 \lesssim \beta\gamma \lesssim 2.75$,
%\begin{align}
% 1.53 \lesssim \beta\gamma \lesssim 2.75,
%\end{align}
hence we can assume $\beta\gamma$ to be $2$. 
%Then, using Eq.\eqref{prob_dec}, 
%the expected number of the decays, $N_{\mathrm{dec}}$, within a distance $l$ is given 
%\begin{align}
% N_{\mathrm{dec}}(l)=N_{\tilde{l}_1} P_{\mathrm{dec}}(l) 
%  = N_{\tilde{l}_1}\left(1-\exp\left(-\frac{l}{\beta\gamma c \tau_{\tilde{l}_1}}\right)\right).
%\end{align}

\begin{table}[t]
\begin{center}
\caption{The expected number of slepton decay in the ATLAS detector. The length is the minimum ($5$ cm) and medium ($50$ cm) 
distance to the pixel detector and the maximum distance to the outer boundary of 
the detectors from the interaction point and corresponds to pixel detector ($3.1$ m), calorimeter ($5.8$ m) 
and muon spectrometer ($25.0$ m). The lifetime is varied from $10^{-11}$ to $10^{-5}$ seconds. 
%The number of sleptons produced at LHC assuming the integrated luminosity, $\mathcal{L}_{\mathrm{int}}=30$ fb$^{-1}$ is $4290$ and 
$\beta\gamma$ is fixed to $2$.}
\begin{tabular}{cccccc}\br
        &~~~$5$ cm~~~ &~~~ $50$ cm~~~ &~~~ $3.1$ m ~~~&~~~ $5.8$ m ~~~&~~~ $25.0$ m ~~~ \\ \mr
~~$10^{-5}$ sec.~~ & $0.04$   & $0.36$   & $2.2$      & $4.1$    & $17.8$ \\ 
$10^{-6}$ sec.     & $0.36$   & $3.6$    & $22.1$     & $41.3$   & $175.1$   \\ 
$10^{-7}$ sec.     & $3.6$    & $35.6$   & $216.0$    & $395.3$  & $1461.9$  \\ 
$10^{-8}$ sec.     & $35.6$   & $343.0$  & $1731.0$   & $2658.3$ & $4223.5$ \\ 
$10^{-9}$ sec.     & $343.0$  & $2425.6$ & $4265.5$   & $4289.7$ & $4290.0$ \\ 
$10^{-10}$ sec.    & $2425.6$ & $4289.0$ & $4290.0$   & $4290.0$ & $4290.0$ \\ 
$10^{-11}$ sec.    & $4289.0$ & $4290.0$ & $4290.0$   & $4290.0$ & $4290.0$ \\ \br
%$10^{-12}$ sec.    & $4290.0$ & $4290.0$ & $4290.0$   & $4290.0$ & $4290.0$ \\ \hline
%$10^{-13}$ sec.    & $4290.0$ & $4290.0$ & $4290.0$    & $4290.0$ & $4290.0$ \\ \hline
\end{tabular}
\label{number_decay}
\end{center}
\end{table}
In Table~\ref{number_decay}, we show the expected number of slepton decays in each detector, assuming the integrated luminosity, 
$\mathcal{L}_{\mathrm{int}}=30$ fb$^{-1}$ and $\beta\gamma=2$. 
As can be seen, when the lifetime, $\tau_{\tilde{l}_1}$, is below $10^{-9}$ sec., most of the sleptons decay inside the pixel detector. 
When $\tau_{\tilde{l}_1} \sim 10^{-8}$ sec., almost half of them decay inside the inner detector and nearly all of them decay within the detector. 
If $\tau_{\tilde{l}_1}$ is between $10^{-8}$ and $10^{-6}$ sec., several hundreds of slepton decays occur inside the ATLAS detector. 
Almost all of them escape from the detector when $\tau_{\tilde{l}_1} > 10^{-5}$ sec., although we expect of the order of $10$ decays 
inside the detector for $\tau_{\tilde{l}_1} \simeq 10^{-5}$ sec.

Sleptons with different lifetimes would give different signatures in the ATLAS detector. 
Depending on the lifetimes, sleptons would decay inside the detector or escape the detector as a stable heavy charged-particle.
For $\tau_{\tilde{l}_1} < 10^{-11}$ sec, since almost all of the sleptons would decay before they reach the first layer of the  
pixel detector, we would not observe any heavy charged-particle track in the detector. 
In this case, it would be more difficult to identify the presence of long-lived sleptons at the ATLAS detector. This
problem will be addressed in a future work \cite{progress}.
For $\tau_{\tilde{l}_1} \sim 10^{-10}$ to $10^{-9}$ sec, almost all the sleptons would decay inside the pixel detector and leave a charged track 
with a kink. Then, a different charged-particle would cross the outer detectors and a corresponding track and/or hit 
would be seen in each detector. Thus, by combining with signals in the outer detectors, we could identify whether the outgoing 
charged-particle is an electron or muon.
Then, if we can fix the mass difference between the lightest slepton and neutralino, we can determine the value of the mass insertion parameters 
from Figs.~\ref{lifetime_right} and \ref{lifetime_left}.
In the case of right-handed slepton mixing case, lifetimes between $10^{-10}$ and $10^{-8}$ sec. would correspond to $(\delta^e_{RR})_{e\tau}$ 
between $10^{-7}$ and $10^{-4}$ with the mass difference, $m_e < \delta m < m_\tau$, and $(\delta^e_{RR})_{\mu\tau}$ between 
$10^{-7}$ and $10^{-5}$ with $ m_\mu < \delta m < m_\tau$. Similarly, in the case of left-handed slepton mixing, the same lifetimes would 
correspond to $(\delta^e_{LL})_{e\tau}$ between $4 \times 10^{-6}$ and $10^{-3}$ with $m_e < \delta m < m_\tau$, 
and $(\delta^e_{LL})_{\mu\tau}$ between $4 \times 10^{-6}$ and $10^{-4}$ with $m_\mu < \delta m < m_\tau$.
For $\tau_{\tilde{l}_1} \sim 10^{-8}$ sec., about half of sleptons would decay inside the inner detector and the rest would decay inside 
the calorimeters and/or muon spectrometer. In this case, it would be important to determine whether the decay is LFV two-body decay 
or lepton flavour conserving three-body decay \cite{progress}. 
For lifetimes between $10^{-7}$ and $10^{-5}$ sec., very few sleptons would decay inside the pixel detector 
and most of them would escape the detector leaving a charged track with a corresponding hit in muon spectrometer. 
Even for particles escaping the detector, we could use muon spectrometer to determine slepton mass and momentum as 
studied in \cite{tarem:2008,tarem:2008ph}.
For $\tau_{\tilde{l}_1} \gtrsim 10^{-5}$ sec., very few sleptons would decay inside the ATLAS detector. In this case, we can put lower 
bound on the lifetime. Then, in a gravity-mediated scenario, this would correspond to a stringent upper bound on $\delta^e$s.

This scenario of long-lived slepton is in principle similar to the situation in gauge-mediated SUSY breaking scenario \cite{Ishiwata:2008tp}. 
For similar lifetimes, more detailed analysis is needed to distinguish scenarios. 
However most of sleptons in gauge-mediated SUSY breaking scenario would escape from the detector and we could see only a few events. 

\section{Summary}
In this work, we have studied lepton flavour violation in a long-lived slepton scenario where the mass difference between 
the NLSP, the lightest slepton, and the LSP, the lightest neutralino, is smaller than the tau mass. With this small mass difference, 
the lifetime of the lightest slepton has a very good sensitivity to small lepton flavour violation parameters 
due to the more than $13$ orders of magnitude suppression of the three-body decay rate with respect to the two-body decay rate. 

We have shown that this small $\delta m$ is possible even in the framework of the CMSSM.
We selected the region in the CMSSM parameter space with $\delta m < m_\tau$, taking into account correct dark matter abundance and direct 
bounds on SUSY and Higgs masses and $b \rightarrow s~\gamma$ constraint.
We have used the experimental results of the anomalous magnetic moment of the muon to indicate the favored regions at given confidence level.
We have found that a small $\delta m$ region consistent with all experimental constraints lies on sizable part of the 
parameter space at $M_{1/2} \gtrsim 700$ GeV, $m_0=100$ to $600$ GeV for different values of $\tan\beta$.
We have also shown that the most favored region, which is consistent with $a_\mu$ at $2~\sigma$, 
corresponds to $240 \lsim m_0 \lsim 260$ GeV and $700 \lsim M_{1/2} \lsim 800$ GeV at $\tan\beta=35$.

Then, we have analyzed the dependence of the lightest slepton lifetimes on different mass insertions, 
$(\delta^e_{RR/LL})_{e\tau,~\mu\tau}$ for values of $\delta m$ from $2.23$ to $0.09$ GeV. 
We found that the lifetimes are proportional to $|\delta^e|^{-2}$ until three- or four-body decays become comparable.
There is a difference of approximately a factor of $10$ in the sensitivity of the lifetime on $\delta^e_{RR}$ and 
$\delta^e_{LL}$. This is due to the different proportion of left-handed and right-handed staus in the lightest slepton.
By comparing the values of the bounds on $\delta^e$s shown in \cite{Kaneko:2008re}, we can see that, in this scenario, 
the lifetimes are sensitive to much smaller values of these $\delta^e$s, even to future sensitivities of proposed experiments. 

Finally, we have discussed the expected phenomenology at LHC experiments, mainly concentrating on the ATLAS detector. 
We have estimated the number of slepton decays in the different detectors, 
assuming an integrated luminosity $\mathcal{L}_{\mathrm{int}}=30$ fb$^{-1}$ and $\beta\gamma=2$ (Table.~\ref{number_decay}).
We have seen that the ATLAS detector can observe lifetimes in the range of $10^{-11}$ to $10^{-6}$ sec., and these lifetime 
would correspond to $(\delta^e_{RR})_{e\tau,\mu\tau}$ between $10^{-7}$ and $10^{-3}$ and $(\delta^e_{LL})_{e\tau,\mu\tau}$ between 
$4 \times 10^{-6}$ and $10^{-3}$.
Therefore we have shown that in the long-lived slepton scenario, the LHC offers a very good opportunity to study lepton 
flavour violation.

\ack
The authors, O.~V. and T.~S,  would like to thank V.~Mitsou for fruitful discussion on detection 
possibilities of the long-lived slepton at the ATLAS detector, and J.~Jones-Perez for numerical 
checks of bounds on mass insertions.
S.K. was supported by European Commission Contracts MRTN-CT-2004-503369 and ILIAS/N6 WP1 
RI I3-CT-2004-506222, and also Funda\c c\~ ao para a Ci\^ encia e a  Tecnologia (FCT, Portugal) 
through CFTP-FCT UNIT 777  which is partially funded through POCTI (FEDER). S.K. is an ER 
supported by the Marie Curie Research Training Network MRTN-CT-2006-035505.
The work of J. S. was supported in part by the Grant-in-Aid for the Ministry of
Education, Culture, Sports, Science, and Technology, Government of Japan Contact Nos.
20025001, 20039001, and 20540251.
The work of T.~S. and O.~V. was supported in part by MEC and FEDER (EC), Grants No. FPA2005-01678 and 
the Generalitat Valenciana for support under the grants PROMETEO/2008/004, GV05/267 and GVPRE/2008/003. 
The work of O.~V was also supported in part by European program MRTN-CT-2006-035482 ``Flavianet''. 
The work of M. Y. was supported in part by the Grant-in-Aid for the Ministry of Education, Culture, Sports, Sc
ience, and Technology, Government of Japan (No. 20007555).


\section*{References}
\begin{thebibliography}{99}
%\cite{Griest:1990kh}
\bibitem{Griest:1990kh}
  Griest K and Seckel D,
  %``Three exceptions in the calculation of relic abundances,''
  {\it Phys. Rev.}  D {\bf 43} (1991) 3191.
  %%CITATION = PHRVA,D43,3191;%%

 %\cite{Jittoh:2005pq}
 \bibitem{Jittoh:2005pq}
  Jittoh T, Sato J, Shimomura T and Yamanaka M,
  %``Long life stau,''
  {\it Phys. Rev.}  D {\bf 73}, 055009 (2006)
  [arXiv:hep-ph/0512197].
  %%CITATION = PHRVA,D73,055009;%%
	 
 %\cite{Jittoh:2007fr}
 \bibitem{Jittoh:2007fr}
  Jittoh J, Kohri K, Koike M, Sato J, Shimomura T and Yamanaka M,
  %``Possible solution to the $^7$Li problem by the long-lived stau,''
  {\it Phys. Rev.}  D {\bf 76}, 125023 (2007)
  [arXiv:0704.2914 [hep-ph]].
  %%CITATION = PHRVA,D76,125023;%%

%\cite{Jittoh:2008eq}
\bibitem{Jittoh:2008eq}
  Jittoh T, Kohri K, Koike M, Sato J, Shimomura T and Yamanaka M,
  %``Stau properties from the Big-Bang Nulceosynthesis and the relic abundance
  %of dark matter,''
  {\it Phys. Rev.}  D {\bf 78}, 055007 (2008)
  [arXiv:0805.3389 [hep-ph]].
  %%CITATION = PHRVA,D78,055007;%%

%\cite{Ryan:2000zz}
\bibitem{Ryan:2000zz}
  Ryan S G, Beers T C, Olive K A, Fields B D and Norris J E,
  %``Primordial Lithium And Big Bang Nucleosynthesis,''
  {\it Astrophys. J.}  {\bf 530} (2000) L57.
  %%CITATION = ASJOA,530,L57;%%

%\cite{Cyburt:2003fe}
\bibitem{Cyburt:2003fe}
  Cyburt R H, Fields B D and Olive K A,
  %``Primordial Nucleosynthesis in Light of WMAP,''
  {\it Phys. Lett.}  B {\bf 567}, 227 (2003)
  [arXiv:astro-ph/0302431].
  %%CITATION = PHLTA,B567,227;%%

%\cite{Coc:2003ce}
\bibitem{Coc:2003ce}
  Coc A, Vangioni-Flam E, Descouvemont P, Adahchour A and Angulo C,
  %``Updated Big Bang Nucleosynthesis confronted to WMAP observations and to the
  %Abundance of Light Elements,''
  {\it Astrophys. J.}  {\bf 600}, 544 (2004)
  [arXiv:astro-ph/0309480].
  %%CITATION = ASJOA,600,544;%%

%\cite{Jarosik:2006ib}
\bibitem{Jarosik:2006ib}
  Jarosik N {\it et al.}  [WMAP Collaboration],
  %``Three-year Wilkinson Microwave Anisotropy Probe (WMAP) observations:  Beam
  %profiles, data processing, radiometer characterization and  systematic error
  %limits,''
  {\it Astrophys. J. Suppl.}  {\bf 170}, 263 (2007)
  [arXiv:astro-ph/0603452].
  %%CITATION = APJSA,170,263;%%

%\cite{Dunkley:2008ie}
\bibitem{Dunkley:2008ie}
  Dunkley J {\it et al.}  [WMAP Collaboration],
  %``Five-Year Wilkinson Microwave Anisotropy Probe (WMAP) Observations:
  %Likelihoods and Parameters from the WMAP data,''
  arXiv:0803.0586 [astro-ph].
  %%CITATION = ARXIV:0803.0586;%%

%\cite{Ross:2000fn}
%\bibitem{Ross:2000fn}
%For a review and further references see:
%\\  G.~G.~Ross,
  %``Models of fermion masses,''
  %\href{http://www.slac.stanford.edu/spires/find/hep/www?irn=4868536}{SPIRES entry}
%{\it Prepared for Theoretical Advanced Study Institute in Elementary Particle Physics (TASI 2000): Flavor Physics 
%for the Millennium, Boulder,
%Colorado, 4-30 Jun 2000}

%\cite{Raidal:2008jk}
%\bibitem{Raidal:2008jk}
%  M.~Raidal {\it et al.},
  %``Flavour physics of leptons and dipole moments,''
%  arXiv:0801.1826 [hep-ph].
  %%CITATION = ARXIV:0801.1826;%%

%\cite{Agashe:1999bm}
\bibitem{Agashe:1999bm}
  Agashe K and Graesser M,
  %``Signals of supersymmetric lepton flavor violation at the LHC,''
  {\it Phys. Rev.}  D {\bf 61} (2000) 075008
  [arXiv:hep-ph/9904422].
  %%CITATION = PHRVA,D61,075008;%%

%\cite{Deppisch:2003wt}
\bibitem{Deppisch:2003wt}
  Deppisch F, Pas H, Redelbach A, Ruckl R and Shimizu Y,
  %``The SUSY seesaw model and lepton-flavor violation at a future electron
  %positron linear collider,''
  {\it Phys. Rev.}  D {\bf 69} (2004) 054014
  [arXiv:hep-ph/0310053].
  %%CITATION = PHRVA,D69,054014;%%

%\cite{Deppisch:2004pc}
\bibitem{Deppisch:2004pc}
  Deppisch F, Kalinowski F, Pas H, Redelbach A and Ruckl R,
  %``Supersymmetric lepton flavour violation at the LHC and LC,''
  arXiv:hep-ph/0401243.
  %%CITATION = HEP-PH/0401243;%%

%\cite{Bartl:2005yy}
\bibitem{Bartl:2005yy}
  Bartl A, Hidaka K, Hohenwarter-Sodek K, Kernreiter T, Majerotto W and Porod W,
	%``Test of lepton flavour violation at LHC,''
  {\it Eur. Phys. J.}  C {\bf 46} (2006) 783
  [arXiv:hep-ph/0510074].
  %%CITATION = EPHJA,C46,783;%%

%\cite{Ibarra:2006sz}
\bibitem{Ibarra:2006sz}
  Ibarra A and Roy S,
  %``Lepton flavour violation in future linear colliders in the long-lived
  %stau NLSP scenario,''
  {\it JHEP} {\bf 0705} (2007) 059
  [arXiv:hep-ph/0606116].
  %%CITATION = JHEPA,0705,059;%%

%\cite{HohenwarterSodek:2007az}
\bibitem{HohenwarterSodek:2007az}
  Hohenwarter-Sodek K and Kernreiter T,
  %``Effects of Lepton Flavour Violation on Chargino Production at the Linear
  %Collider,''
  {\it JHEP} {\bf 0706} (2007) 071
  [arXiv:0704.2684 [hep-ph]].
  %%CITATION = JHEPA,0706,071;%%


%\cite{Hirsch:2008dy}
\bibitem{Hirsch:2008dy}
  Hirsch M, Valle J W F, Porod W, Romao J C and Villanova del Moral A,
  %``Probing minimal supergravity in type I seesaw with lepton flavour violation
  %at the LHC,''
  arXiv:0804.4072 [hep-ph].
  %%CITATION = ARXIV:0804.4072;%%


%\cite{Hirsch:2008gh}
\bibitem{Hirsch:2008gh}
  Hirsch M, Kaneko S and Porod W,
  %``Supersymmetric seesaw type II: LHC and lepton flavour violating
  %phenomenology,''
  arXiv:0806.3361 [hep-ph].
  %%CITATION = ARXIV:0806.3361;%%


%\cite{Hamaguchi:2006vu}
\bibitem{Hamaguchi:2006vu}
  Hamaguchi K, Nojiri M M and de Roeck A,
  %``Prospects to study a long-lived charged next lightest supersymmetric
  %particle at the LHC,''
  {\it JHEP} {\bf 0703}, 046 (2007)
  [arXiv:hep-ph/0612060].
  %%CITATION = JHEPA,0703,046;%%

%\cite{tarem:2008}
\bibitem{tarem:2008}
  Tarem S {\it et al.} [ATLAS Collaboration]
  %``Trigger and Reconstruction for heavy long-lived charged particles with the ATLAS detector''
  ATL-SN-ATLAS-2008-071

%\cite{tarem:2008ph}
\bibitem{tarem:2008ph}
  Tarem S {\it et al.}
  %``Trigger and Reconstruction for heavy long-lived charged particles with the ATLAS detector''
  ATL-PHYS-PUB-2005-02

%\cite{Ishiwata:2008tp}
\bibitem{Ishiwata:2008tp}
  Ishiwata K, Ito T and Moroi T,
  %``Long-Lived Unstable Superparticles at the LHC,''
  arXiv:0807.0975 [hep-ph].
  %%CITATION = ARXIV:0807.0975;%%

%\cite{Porod:2003um}
\bibitem{Porod:2003um}
  Porod W,
  %``SPheno, a program for calculating supersymmetric spectra, SUSY particle
  %decays and SUSY particle production at e+ e- colliders,''
  {\it Comput. Phys. Commun.}  {\bf 153}, 275 (2003)
  [arXiv:hep-ph/0301101].
  %%CITATION = CPHCB,153,275;%%

%\cite{Belanger:2008sj}
\bibitem{Belanger:2008sj}
  Belanger G, Boudjema F, Pukhov A and Semenov A,
  %``Dark matter direct detection rate in a generic model with micrOMEGAs2.1,''
  arXiv:0803.2360 [hep-ph].
  %%CITATION = ARXIV:0803.2360;%%

%\cite{Bennett:2006fi}
\bibitem{Bennett:2006fi}
  Bennett G W {\it et al.}  [Muon G-2 Collaboration],
  %``Final report of the muon E821 anomalous magnetic moment measurement at
  %BNL,''
  {\it Phys. Rev.}  D {\bf 73}, 072003 (2006)
  [arXiv:hep-ex/0602035].
  %%CITATION = PHRVA,D73,072003;%%

%\cite{Davier:2007ua}
\bibitem{Davier:2007ua}
  Davier M,
  %``The hadronic contribution to (g-2)(mu),''
  {\it Nucl. Phys. Proc. Suppl.}  {\bf 169}, 288 (2007)
  [arXiv:hep-ph/0701163].
  %%CITATION = NUPHZ,169,288;%%

%\cite{Ellis:1999mm}
\bibitem{Ellis:1999mm}
  Ellis J R, Falk T, Olive K A and Srednicki M,
  %``Calculations of neutralino stau coannihilation channels and the
  %cosmologically relevant region of MSSM parameter space,''
  {\it Astropart. Phys.}  {\bf 13} (2000) 181
  [Erratum-ibid.\  {\bf 15} (2001) 413]
  [arXiv:hep-ph/9905481].
  %%CITATION = APHYE,13,181;%%	 

%\cite{Kaneko:2008re}
\bibitem{Kaneko:2008re}
  Kaneko S, Sato J, Shimomura T, Vives O and Yamanaka M,
  %``Measuring Lepton Flavour Violation at LHC with Long-Lived Slepton in the
  %Coannihilation Region,''
  {\it Phys. Rev.}  D {\bf 78}, 116013 (2008)
  [arXiv:0811.0703 [hep-ph]].
  %%CITATION = PHRVA,D78,116013;%%

%\bibitem{Canadas:2008}
%	B.~Ca\~nadas, D.G.~Cerde\~no, C.~Mu\~noz and S.~Panda, work in progress.

% \bibitem{Hikasa_note}
%	K.~Hikasa and his friends, ``Minimal Supersymmetry for Collider Physicists''. 
	 
 %\cite{Ciuchini:2007ha}
% \bibitem{Ciuchini:2007ha}
%  M.~Ciuchini, A.~Masiero, P.~Paradisi, L.~Silvestrini, S.~K.~Vempati and O.~Vives,
  %``Soft SUSY breaking grand unification: Leptons vs quarks on the flavor
  %playground,''
%  Nucl.\ Phys.\  B {\bf 783}, 112 (2007)
%  [arXiv:hep-ph/0702144].
  %%CITATION = NUPHA,B783,112;%%

%%\cite{Borzumati:1986qx}
%\bibitem{Borzumati:1986qx}
%  F.~Borzumati and A.~Masiero,
%  %``Large Muon And Electron Number Violations In Supergravity Theories,''
%  Phys.\ Rev.\ Lett.\  {\bf 57}, 961 (1986).
%  %%CITATION = PRLTA,57,961;%%
%
%%\cite{Hisano:1995nq}
%\bibitem{Hisano:1995nq}
%  J.~Hisano, T.~Moroi, K.~Tobe, M.~Yamaguchi and T.~Yanagida,
%  % ``Lepton flavor violation in the supersymmetric standard model with seesaw
%  %induced neutrino masses,''
%  Phys.\ Lett.\  B {\bf 357}, 579 (1995)
%  [arXiv:hep-ph/9501407].
%  %%CITATION = PHLTA,B357,579;%%
%
%%\cite{Hisano:1998fj}
%\bibitem{Hisano:1998fj}
%  J.~Hisano and D.~Nomura,
%  % ``Solar and atmospheric neutrino oscillations and lepton flavor violation  in
%  %supersymmetric models with the right-handed neutrinos,''
%  Phys.\ Rev.\  D {\bf 59}, 116005 (1999)
%  [arXiv:hep-ph/9810479].
%  %%CITATION = PHRVA,D59,116005;%%
%
%%\cite{Sato:2000ff}
%\bibitem{Sato:2000ff}
%  J.~Sato and K.~Tobe,
%  %``Neutrino masses and lepton-flavor violation in supersymmetric models  with
%  %lopsided Froggatt-Nielsen charges,''
%  Phys.\ Rev.\  D {\bf 63}, 116010 (2001)
%  [arXiv:hep-ph/0012333].
%  %%CITATION = PHRVA,D63,116010;%%
%
%%\cite{Sato:2000zh}
%\bibitem{Sato:2000zh}
%  J.~Sato, K.~Tobe and T.~Yanagida,
%  %``A constraint on Yukawa-coupling unification from lepton-flavor  violating
%  %processes,''
%  Phys.\ Lett.\  B {\bf 498}, 189 (2001)
%  [arXiv:hep-ph/0010348].
%  %%CITATION = PHLTA,B498,189;%% 
%
%%\cite{Casas:2001sr}
%\bibitem{Casas:2001sr}
%  J.~A.~Casas and A.~Ibarra,
%  %``Oscillating neutrinos and mu --> e, gamma,''
%  Nucl.\ Phys.\  B {\bf 618} (2001) 171
%  [arXiv:hep-ph/0103065].
%  %%CITATION = NUPHA,B618,171;%%
%
%%\cite{Lavignac:2001vp}
%\bibitem{Lavignac:2001vp}
%  S.~Lavignac, I.~Masina and C.~A.~Savoy,
%  %``tau --> mu gamma and mu --> e gamma as probes of neutrino mass models,''
%  Phys.\ Lett.\  B {\bf 520} (2001) 269
%  [arXiv:hep-ph/0106245].
%  %%CITATION = PHLTA,B520,269;%%
%
%%\cite{Kageyama:2001zt}
%\bibitem{Kageyama:2001zt}
%  A.~Kageyama, S.~Kaneko, N.~Shimoyama and M.~Tanimoto,
%  %``Lepton flavor violating process in degenerate and inverse-hierarchical
%  %neutrino models,''
%  Phys.\ Lett.\  B {\bf 527}, 206 (2002)
%  [arXiv:hep-ph/0110283].
%  %%CITATION = PHLTA,B527,206;%%
%
%%\cite{Kageyama:2001tn}
%\bibitem{Kageyama:2001tn}
%  A.~Kageyama, S.~Kaneko, N.~Shimoyama and M.~Tanimoto,
%  %``Lepton flavor violating processes in bi-maximal texture of neutrino
%  %mixings,''
%  Phys.\ Rev.\  D {\bf 65}, 096010 (2002)
%  [arXiv:hep-ph/0112359].
%  %%CITATION = PHRVA,D65,096010;%%
%
%%\cite{Masiero:2002jn}
%\bibitem{Masiero:2002jn}
%  A.~Masiero, S.~K.~Vempati and O.~Vives,
%  %``Seesaw and lepton flavour violation in SUSY SO(10),''
%  Nucl.\ Phys.\  B {\bf 649}, 189 (2003)
%  [arXiv:hep-ph/0209303].
%  %%CITATION = NUPHA,B649,189;%%
%
%%\cite{Deppisch:2002vz}
%\bibitem{Deppisch:2002vz}
%  F.~Deppisch, H.~Pas, A.~Redelbach, R.~Ruckl and Y.~Shimizu,
%  %``Probing the Majorana mass scale of right-handed neutrinos in mSUGRA,''
%  Eur.\ Phys.\ J.\  C {\bf 28} (2003) 365
%  [arXiv:hep-ph/0206122].
%  %%CITATION = EPHJA,C28,365;%%
%
%%\cite{Masiero:2004js}
%\bibitem{Masiero:2004js}
%  A.~Masiero, S.~K.~Vempati and O.~Vives,
%  %``Massive neutrinos and flavour violation,''
%  New J.\ Phys.\  {\bf 6} (2004) 202
%  [arXiv:hep-ph/0407325].
%  %%CITATION = NJOPF,6,202;%%
%
%%\cite{Ota:2005et}
%\bibitem{Ota:2005et}
%  T.~Ota and J.~Sato,
%  %``Signature of the minimal supersymmetric standard model with  right-handed
%  %neutrinos in long baseline experiments,''
%  Phys.\ Rev.\  D {\bf 71}, 096004 (2005)
%  [arXiv:hep-ph/0502124].
%  %%CITATION = PHRVA,D71,096004;%%
%
%%\cite{Petcov:2005jh}
%\bibitem{Petcov:2005jh}
%  S.~T.~Petcov, W.~Rodejohann, T.~Shindou and Y.~Takanishi,
%  %``The see-saw mechanism, neutrino Yukawa couplings, LFV decays l(i) -->  l(j)
%  %+ gamma and leptogenesis,''
%  Nucl.\ Phys.\  B {\bf 739} (2006) 208
%  [arXiv:hep-ph/0510404].
%  %%CITATION = NUPHA,B739,208;%%
%
%%\cite{Antusch:2006vw}
%\bibitem{Antusch:2006vw}
%  S.~Antusch, E.~Arganda, M.~J.~Herrero and A.~M.~Teixeira,
%  %``Impact of theta(13) on lepton flavour violating processes within SUSY
%  %seesaw,''
%  JHEP {\bf 0611} (2006) 090
%  [arXiv:hep-ph/0607263].
%  %%CITATION = JHEPA,0611,090;%%
%
%%\cite{Ross:2004qn}
%\bibitem{Ross:2004qn}
%  G.~G.~Ross, L.~Velasco-Sevilla and O.~Vives,
%  %``Spontaneous CP violation and non-Abelian family symmetry in SUSY,''
%  Nucl.\ Phys.\  B {\bf 692} (2004) 50
%  [arXiv:hep-ph/0401064].
%  %%CITATION = NUPHA,B692,50;%%
%
%%\cite{Antusch:2008jf}
%\bibitem{Antusch:2008jf}
%  S.~Antusch, S.~F.~King, M.~Malinsky and G.~G.~Ross,
%  %``Solving the SUSY Flavour and CP Problems with Non-Abelian Family Symmetry
%  %and Supergravity,''
%  arXiv:0807.5047 [hep-ph].
%  %%CITATION = ARXIV:0807.5047;%%
%
%%\cite{Calibbi:2008qt}
%\bibitem{Calibbi:2008qt}
%  L.~Calibbi, J.~Jones-Perez and O.~Vives,
%  %``Electric dipole moments from flavoured CP violation in SUSY,''
%  arXiv:0804.4620 [hep-ph].
%  %%CITATION = ARXIV:0804.4620;%%
%
%%\cite{Feruglio:2008ht}
%\bibitem{Feruglio:2008ht}
%  F.~Feruglio, C.~Hagedorn, Y.~Lin and L.~Merlo,
%  %``Lepton Flavour Violation in Models with A4 Flavour Symmetry,''
%  arXiv:0807.3160 [hep-ph].
%  %%CITATION = ARXIV:0807.3160;%%
%
%%\cite{Masina:2002mv}
\bibitem{Masina:2002mv}
  Masina I and Savoy C A,
%  % ``Sleptonarium (constraints on the CP and flavour pattern of scalar lepton
%  %masses),''
  {\it Nucl. Phys.}  B {\bf 661}, 365 (2003)
  [arXiv:hep-ph/0211283].
%  %%CITATION = NUPHA,B661,365;%%
%
%%\cite{Barbieri:1994pv}
%\bibitem{Barbieri:1994pv}
%  R.~Barbieri and L.~J.~Hall,
%  %``Signals for supersymmetric unification,''
%  Phys.\ Lett.\  B {\bf 338} (1994) 212
%  [arXiv:hep-ph/9408406].
%  %%CITATION = PHLTA,B338,212;%%
%
%%\cite{Barbieri:1995tw}
%\bibitem{Barbieri:1995tw}
%  R.~Barbieri, L.~J.~Hall and A.~Strumia,
%  %``Violations of lepton flavor and CP in supersymmetric unified theories,''
%  Nucl.\ Phys.\  B {\bf 445} (1995) 219
%  [arXiv:hep-ph/9501334].
%  %%CITATION = NUPHA,B445,219;%%
%
%%\cite{Calibbi:2006ne}
%\bibitem{Calibbi:2006ne}
%  L.~Calibbi, A.~Faccia, A.~Masiero and S.~K.~Vempati,
%  %``Running $U_{e3}$ and BR($\mu\to e +\gamma$) in SUSY-GUTs,''
%  JHEP {\bf 0707} (2007) 012
%  [arXiv:hep-ph/0610241].
%  %%CITATION = JHEPA,0707,012;%%
%
%%\cite{Calibbi:2007bk}
%\bibitem{Calibbi:2007bk}
%  L.~Calibbi, Y.~Mambrini and S.~K.~Vempati,
%  %``SUSY-GUTs, SUSY-Seesaw and the Neutralino Dark Matter,''
%  JHEP {\bf 0709} (2007) 081
%  [arXiv:0704.3518 [hep-ph]].
%  %%CITATION = JHEPA,0709,081;%%
%

%\cite{:2008zz}
\bibitem{:2008zz}
  Bentvelsen S {\it et al.}  [ATLAS Collaboration],
  %``The ATLAS Experiment at the CERN Large Hadron Collider,''
  {\it JINST} {\bf 3} (2008) S08003
  %%CITATION = JINST,3,S08003;%%

\bibitem{atlashp_url}
  For more information on the ATLAS Exepriment, visit website~http://atlas.ch/

%\cite{:2008zzk}
\bibitem{:2008zzk}
  Adolphi A {\it et al.}  [CMS Collaboration],
  %``The CMS experiment at the CERN LHC,''
  {\it JINST} {\bf 3} (2008) S08004
  %%CITATION = JINST,3,S08004;%%

%\cite{Battaglia:2001zp}
\bibitem{Battaglia:2001zp}
  Battaglia M {\it et al.},
  %``Proposed post-LEP benchmarks for supersymmetry,''
  {\it Eur. Phys. J.}  C {\bf 22}, 535 (2001)
  [arXiv:hep-ph/0106204].
  %%CITATION = EPHJA,C22,535;%%

%\cite{Battaglia:2003ab}
\bibitem{Battaglia:2003ab}
  Battaglia M, De Roeck A, Ellis J R, Gianotti F, Olive K A and Pape L,
  %``Updated post-WMAP benchmarks for supersymmetry,''
  {\it Eur. Phys. J.}  C {\bf 33}, 273 (2004)
  [arXiv:hep-ph/0306219].
  %%CITATION = EPHJA,C33,273;%%

\bibitem{diff_tanb}
	For different values of $\tan\beta$, branching ratios for each decay are different but total branching ratio will 
	be the same.

%\cite{Skands:2001it}
\bibitem{Skands:2001it}
  Skands P Z,
  %``Searching for L-violating supersymmetry at the LHC,''
  {\it Eur. Phys. J.}  C {\bf 23}, 173 (2002)
  [arXiv:hep-ph/0110137].
  %%CITATION = EPHJA,C23,173;%%

\bibitem{progress}
  Kaneko S, Sato J, Shimomura T, Vives O and Yamanaka M, work in progress.

\end{thebibliography}
\end{document}